\documentclass[10pt,pre,byrevtex,superscriptaddress,bibnotes]{revtex4}
\usepackage{graphics}

\begin{document}

\begin{center}\textit{Published in}
\textbf{ Europhysics News 37/2 (2006) 10}                                                      \end{center}
\bigskip

\title{Reply to  the Comment by T. Dauxois, F. Bouchet, S. Ruffo on the paper \\
by A. Rapisarda and A. Pluchino, Europhysics News, 36 (2005)  202}
\author{ Andrea Rapisarda and Alessandro Pluchino}
\affiliation{Dipartimento di Fisica e Astronomia and Infn Universit\'a  di Catania, 
Via S. Sofia 64, I-95123, Catania, Italy\\
e-mail:
\texttt{ andrea.rapisarda@ct.infn.it ~~~
alessandro.pluchino@ct.infn.it}}
\bigskip
\date{\today}
\maketitle


In the comment by T.Dauxois  et al.  \cite{comment} the authors question our application of nonextensive statistical mechanics  proposed by Tsallis \cite{tsa1} and discussed in \cite{epnhmf1} to explain
the anomalous dynamics of the Hamiltonian Mean Field (HMF) model.  
More specifically they  claim that the explanation of the metastability 
found in the out-of-equilibrium dynamics is only a  fitting procedure and is also  in contrast with a previous application done in ref. \cite{pre1}. This criticism mostly relies on recent studies based on the Vlasov approach and discussed in refs.  \cite{bouchet,yama}, where the authors claim to explain  the anomalous  behaviour of the HMF model in terms of a  standard  formalism.
In order to reply to this comment we want to stress  a few numerical facts and conclude with some final considerations. 

{\it (i)} In our numerical simulations we consider always a {\it finite} number of particles, which plays the role of a collision term  absent in the  Vlasov equation.  This collision term is very important since it drives the systems  towards the equilibrium. 

{\it (ii)} In our paper \cite{epnhmf1}, we use  a {\it finite} initial  magnetization which leads to a  violent thermal explosion.  The   quasi-stationary state which follows is microscopically nonhomogeneous, with a hierarchical cluster size distribution \cite{hmfvetri}. The Vlasov-like  approach  proposed in \cite{bouchet} has severe problems in dealing with these inhomogeneities. 
Up to now all the derivations presented in the literature start from a homogenous metastable state, where no violent relaxation occurs \cite{cretahmf1}. In this case, the 
decay of the  velocity correlation function is very fast (almost exponential), in remarkable contrast to what observed for an intial finite magnetization, where a q-exponential (with $q>1$) is found \cite{epnhmf1,cretahmf1}.

{\it (iii)} The predictions of  Tsallis thermostatistics \cite{tsa1}   are successfully compared with the numerical results, as shown in figs.3,4 of ref. \cite{epnhmf1}. In this case, it is not true that we perform simply a  fit of numerical data. By means of Tsallis statistics and using q-exponentials to reproduce extremely well  the anomalous diffusion behaviour, we can predict the correlation decay with great precision and viceversa. At variance,  the results of the approach proposed by Dauxois et al. \cite{bouchet} have not been tested with numerical simulations, so that no real prediction can reasonably be claimed. 

{\it (iv)} The results presented in \cite{epnhmf1} are {\it not} in contradiction with previous papers since they refer to velocity correlations decay and {\it not} to the marginal velocity probability density functions discussed in \cite{pre1},  where the  entropic index  extracted was only an effective one and indicated a strong departure from a Gaussian shape. On the other hand, the possible application of Tsallis statistics in long-range Hamiltonian systems is confirmed by several other studies \cite{epnhmf2}.

In conclusion the HMF model is a paradigmatic example of a large class of long-range Hamiltonian systems  which have important  physical applications, ranging from self-gravitating systems to plasmas.
The nonhomogeneous metastability observed for  the HMF model  goes {\it undoubtedly} beyond standard Boltzmann-Gibbs statistical mechanics and has a dynamical origin,  therefore a new kind of kinetics should be used \cite{chava2}.
In general, adopting different perspectives is a useful procedure to shed light on a tricky problem. Tsallis statistics is a good candidate to explain and interpret the strange behaviour of long-range Hamiltonian systems, and this  is {\it not} in contradiction with other possible formalisms, including that one of Dauxois et al. (analogously the Langevin and the Fokker-Planck phenomenological formulations   are {\it not} in contradiction with Boltzmann-Gibbs statistical mechanics). We have also successfully applied techniques normally used for glassy systems \cite{hmfvetri}, and interesting connections with Kuramoto model and the synchronization problem have been advanced \cite{cretahmf1}.  In any case further work is needed to understand in detail this intriguing   new field.


\end{document}